\documentstyle[prl,aps,multicol]{revtex}
\begin{document}
\author{K.L. Sebastian}
\address{Department of Inorganic and Physical Chemistry\\
Indian Institute of Science\\
Bangalore 560012\\
India}
\title{Molecular ratchets - verification of the principle of detailed balance and
driving them in one direction}
\maketitle

\begin{abstract}
We argue that the recent experiments of Kelly et. al.(Angew. Chem. Int. Ed.
Engl. 36, 1866 (1997)) on molecular ratchets, in addition to being in
agreement with the second law of thermodynamics, is a test of the principle
of detailed balance for the ratchet. We suggest new experiments, using an
asymmetric ratchet, to further test the principle. We also point out methods
involving a time variation of the temperature to to give it a directional
motion
\end{abstract}

It was pointed out long ago by Feynman \cite{Feynman} that a microscopic
ratchet, in equilibrium with an isothermal heat bath cannot have a net
rotation in any direction - otherwise, the ratchet can be used to extract
work from an isothermal system, which is a violation of the second law of
thermodynamics. Recently, in a very interesting paper, Kelly et. al.\cite
{Kelly} reported the synthesis and the study of the rotational motion of a
molecular ratchet. They found the rotation of the ratchet to occur with
equal likelihood in either direction, and they conclude that this is in
agreement with the second law of thermodynamics (see also comment on this
paper by Davis\cite{Davis})

In the following, we argue that the experiment not only verifies the second
law of thermodynamics, but it also provides a direct test of the principle
of detailed balance. Our argument is based upon the fact that the experiment
is equivalent to putting a label on the Hydrogens which are opposite the
pawl and then probing their dynamics under the rotation of the ratchet. By
putting such a label, we are preparing the system in a rather special, but
non-equilibrium state (see below). As time passes, the probability
distribution evolves and eventually would reach equilibrium. Hence the fact
that results of the experiment show no net rotation is surprising! We argue
that this results from detailed balance and hence in this experiment, one is
verifying more than the second law - actually the principle of detailed
balance. We suggest new experiments involving an asymmetric ratchet which
would further prove this conclusion. We also suggest a way to cause the
symmetric ratchet to undergo a net directional motion, which should be
possible to experimentally observe.

In the experiment, first the spin of the atom H$_a$ in the molecule is
selectively polarized. This means that a population inversion of the spin
states of these atoms has been caused. Then, as the internal rotation
proceeds, H$_a$ gets converted into H$_b$ or H$_c$ depending on the
direction in which the rotation happens, resulting in a transfer of the
polarization and the amount of this transfer is measured.

We denote the population difference between the up and down states of H$_a$
at the time $t$ by $\Delta N_a(t)$. Its equilibrium value is $\Delta
N_{a,e}=-N_0\frac{1-\mu }{1+\mu }$ , where $N_0$ is the total number of
molecules and $\mu =e^{-\Delta E/(k_BT)}$, $\Delta E$ being the energy
difference between the up and the down spin states. Let $n_{{\cal A}%
}(t)=\Delta N_a(t)-\Delta N_{a,e}$ denote the deviation of $\Delta N_a(t)$
from its equilibrium value. Its initial value is $n_{{\cal A}}(0)=2N_0\frac{%
1-\mu }{1+\mu }$.

The molecular ratchet can undergo internal rotation and the corresponding
angle coordinate is denoted by $\varphi $. It varies in the range $\left(
-\pi ,\pi \right) $. We divide this range in to three regions ${\cal A}$ $%
\equiv $ $\left( -\pi /3,\pi /3\right) $, ${\cal B}$ $\equiv \left( \pi
/3,\pi \right) $ and ${\cal C}\equiv \left( -\pi ,-\pi /3\right) $ (see the
figure 1). The equilibrium probability distribution $P_e(\varphi )$ (see
below) is shown in the figure 2(a). At equilibrium, all the three regions
are equally likely. When H$_a$ is selectively spin polarized, one is
effectively putting a label on a population $n_{{\cal A}}(0)$ of the
molecules, which have $\varphi $ in the range ${\cal A}$. The experiment
studies the dynamics of internal rotation of these molecules by measuring
the amounts $n_{{\cal B}}(t)$ and $n_{{\cal C}}(t)$ crossing over to the
other regions ${\cal B}$ and ${\cal C}$. The rotational motion may be taken
to obey the diffusion equation

\begin{equation}
\label{one}\frac{\partial P(\varphi ,t)}{\partial t}=\left\{ \frac \partial {%
\partial \varphi }V^{\prime }(\varphi )+k_BT\frac{\partial ^2}{\partial
\varphi ^2}\right\} P(\varphi ,t) 
\end{equation}
We have absorbed the (unnecessary) constants into our definitions of
variables because of which the ''time'' $t$ has now dimensions of 1/energy. $%
V(\varphi )$ is the potential energy for the (internal) rotation. It has an
asymmetric form making the molecule a ratchet \cite{Kelly}. We shall neglect
spin relaxation in our analysis. The above equation has an equilibrium state
with $P_e(\varphi )={\cal N}e^{-\beta V(\varphi )}$, where ${\cal N=}%
1/\int_{-\pi }^\pi d\varphi e^{-\beta V(\varphi )}$ with $\beta =1/(k_BT)$.
As $V(\varphi )$ is periodic with period $2\pi /3$, the equilibrium
probability distribution too is periodic with the same period.

The spin polarization of H$_a$ is due to an initial distribution with the
excess population spread {\bf only over the region } ${\cal A}$ with a
probability distribution $P_e(\varphi )$. That is, $P(\varphi
,0)=3P_e(\varphi )$ if $-\pi /3<\varphi <\pi /3$ and $P(\varphi ,0)=0$
otherwise (The numerical factor 3 is put to ensure normalization. The number
density of molecules in the population, having an angle $\varphi $ is $n_{%
{\cal A}}(0)P(\varphi ,0)$). To calculate the values of $n_{{\cal A}}(t)$, $%
n_{{\cal B}}(t)$ and $n_{{\cal C}}(t)$, we need to look at the dynamics of
this population. For this, we have to solve the equation (\ref{one}) subject
to this initial condition and then calculate $n_{{\cal I}}(t)=\int_Id\varphi
P(\varphi ,t)$, for ${\cal I=A,B,C}$. This initial probability distribution
function is shown by the full line in figure 2(b).

The initial probability distribution $P(\varphi ,0)$ is a truncated
equilibrium probability function, truncated to zero outside the region $%
{\cal A}$. The second law and the symmetry of the ratchet requires that the
amounts that pass over to ${\cal B}$ and ${\cal C}$ would be the same
initially - that is, at $t=0$, $\frac{dn_{{\cal B}}(t)}{dt}=\frac{dn_{{\cal C%
}}(t)}{dt}$ . However, as time passes, one expects $P(\varphi ,t)$ to become
a truly non-equilibrium probability distribution (a typical one is shown by
the dotted curve of figure 2(b)), and hence one would expect that $n_{_{%
{\cal B}}}(t)\neq n_{{\cal C}}(t)$, in general, even though, experiment
shows the two are equal. We now ask why is this so.

The solution of the equation (\ref{one}) may be written as

\begin{equation}
\label{three}P(\varphi ,t)=\int_{-\pi }^\pi d\varphi _1G(\varphi ,t;\varphi
_1,0)P(\varphi _1,0) 
\end{equation}
where $G(\varphi ,t;\varphi _1,0)$ is the Green's function for the
differential equation in (\ref{one}). The principle of detailed balance
implies \cite{Kampen}

\begin{equation}
\label{four}G(\varphi ,t;\varphi _1,0)P_e(\varphi _1)=G(\varphi _1,t;\varphi
,0)P_e(\varphi ) 
\end{equation}

It is easy to derive this equation starting from the equation (\ref{one}).
The equation (\ref{three}) can be written as

$$
P(\varphi ,t)=3\int_{{\cal A}}d\varphi _1G(\varphi ,t;\varphi
_1,0)P_e(\varphi _1) 
$$

Now, $n_{{\cal B}}(t)=\int_{{\cal B}}d\varphi P(\varphi ,t)$ $=3\int_{{\cal B%
}}d\varphi \int_{{\cal A}}d\varphi _1G(\varphi ,t;\varphi _1,0)P_e(\varphi
_1)$. Using the detailed balance condition of equation (\ref{four}) we get

\begin{equation}
\label{five}n_{{\cal B}}(t)=3\int_{{\cal A}}d\varphi \int_{{\cal B}}d\varphi
_1G(\varphi ,t;\varphi _1,0)P_e(\varphi _1). 
\end{equation}

As the potential is a periodic function, with period $2\pi /3$, the
propagator and the equilibrium probability distribution too are periodic
functions with the same period of $2\pi /3$. Hence we can write

\begin{equation}
\label{six}n_{{\cal B}}(t)=3\int_{{\cal C}}d\varphi \int_{{\cal A}}d\varphi
_1G(\varphi ,t;\varphi _1,0)P_e(\varphi _1) 
\end{equation}
$$
=\int_{{\cal C}}d\varphi P(\varphi ,t) 
$$

\begin{equation}
\label{seven}=n_{{\cal C}}(t) 
\end{equation}

Thus, though the probability distribution would develop in to a
non-equilibrium one as in figure 2(b), the distribution is rather special
and $n_{{\cal B}}(t)$ $=n_{{\cal C}}(t)$ at all times! Further, it is also
clear that one can arrive at the same conclusion for any problem for which
the equations (\ref{three}) and (\ref{four}) are valid. Having proved the
general result, we ask: how can one overcome this, and cause $n_{{\cal B}%
}(t) $ $\neq n_{{\cal C}}(t)$? Noticing that our arguments made use of the
periodicity of the potential $V(\varphi )$, we conclude that if one had an
asymmetric ratchet, like the one in the figure 3, the step from equation (%
\ref{five}) to (\ref{six}) would not go through.

Hence $n_{{\cal B}}(t)$ cannot be equal to $n_{{\cal C}}(t)$, and this
should be seen if an experiment similar to that of Kelly et. al is
performed. Making the ratchet asymmetric is not difficult - one would have
to use a molecule like the one in the figure 4. It is also possible to use
such a molecule for a more stringent test of the principle of detailed
balance. One first polarizes H$_a$ and measures $n_{{\cal B}}(t)$ and then
polarizes H$_b$ and then measures $n_{{\cal A}}(t)$ - detailed balance
implies that the two have to be equal. A similar test can be done with the
molecule of Kelly too (though it has not been done), but an experiment with
an asymmetric ratchet would be more interesting. An easy experiment to make
the molecule have a net transient motion in one direction is to have a
sudden temperature jump in the experiments of Kelly et. al. immediately
after spin polarizing H$_a$. This should lead to $n_{{\cal B}}(t)\neq n_{%
{\cal C}}(t)$ which can then be experimentally observed. Finally, it is
possible to vary the temperature periodically in time - this would
correspond to a Carnot cycle for the molecular ratchet. This will cause the
system to settle into a steady state with net rotation in one direction. We
have performed model calculations and computer simulations and verified
these possibilities \cite{KLS}. In principle, when ultrasonic waves pass
through a liquid containing the molecular ratchet, transfer of energy to the
rotational motion of the ratchet, from the translational motion of the
surrounding liquid molecules can set the ratchet in a steady state with net
rotation in one direction.

I thank Professors E. Arunan, J. Chandrasekhar, S. Ramakrishnan and S.K.
Rangarajan and A. Chakraborty for interesting discussions.

{\bf Figure Captions}

\begin{enumerate}
\item  Figure 1: The ratchet and the regions ${\cal A}$, ${\cal B}$ and $%
{\cal C}$ 

\item  Figure 2: (a) The equilibrium probability distribution against the
angle co-ordinate. (b) The full line shows the initial probability
distribution. It develops into a non-equilibrium distribution of the type
shown by the dotted line.

\item   Figure 3: The asymmetric ratchet. Notice that the teeth are of
different sizes. 

\item  Figure 4: An asymmetric molecular ratchet.
\end{enumerate}

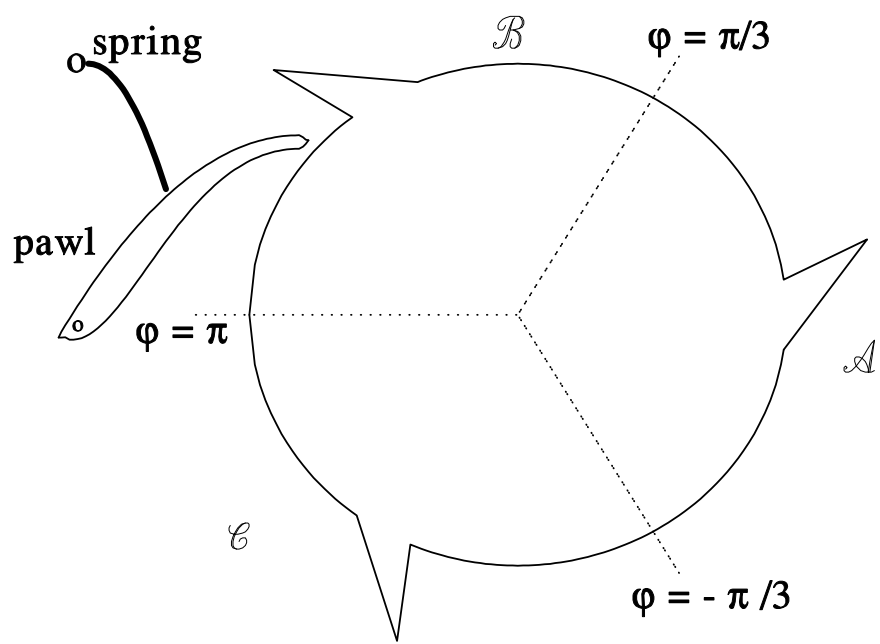
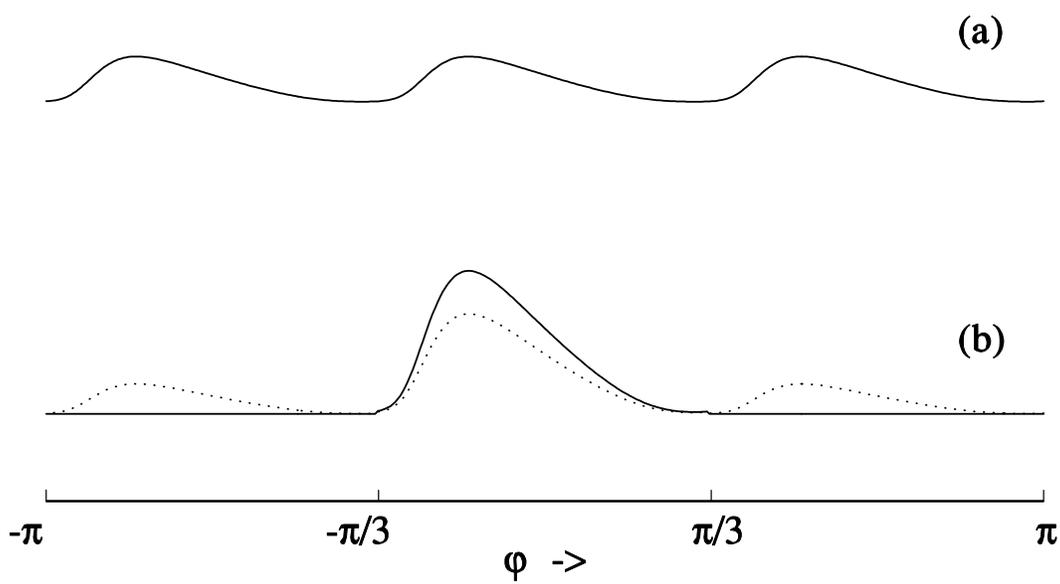
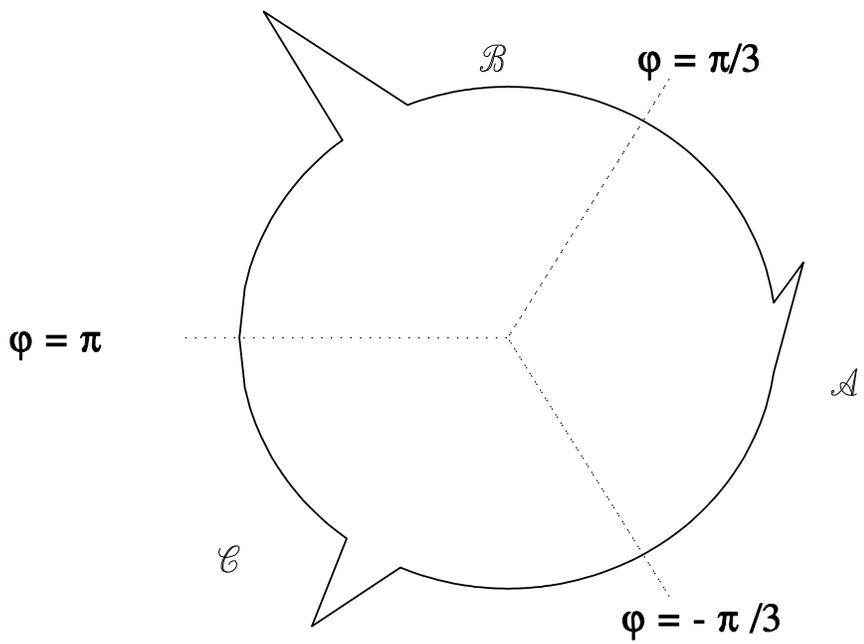
\end{document}